\documentclass[twocolumn,showpacs,preprintnumbers,amsmath,amssymb]{revtex4-1}

\usepackage[abs]{overpic}

\usepackage{dcolumn}
\usepackage{bm}
\usepackage{amsmath,amssymb,amsthm,bm,helvet}
\usepackage[mathscr]{eucal}
\usepackage{graphicx}
\usepackage{tikz}

\newcommand{\beq}{\begin{equation}}
\newcommand{\eeq}{\end{equation}}

\newcommand{\bff}{\begin{figure}[h]\begin{center}}
\newcommand{\eff}{\end{center}\end{figure}}
\newcommand{\ai}{\left\langle 1/X\right\rangle}

\begin{document}

\title{A  gas of elongated objects; an analytical approach}

\author{Mohammad H. Ansari}
\email{mhansari@uwaterloo.ca}
\affiliation{Department of Combinatorics and Optimizations, Faculty of Mathematics, University of Waterloo, Waterloo, ON, Canada}
\affiliation{Institute for Quantum Computing and Department of Physics and Astronomy,  University of Waterloo, Waterloo, ON, Canada}
\date{\today}

\begin{abstract}
We calculate a collective number of thermodynamic quantities in a one-dimensional gas of hard elongated objects (such as needles) whose centers  mobile  on a line. Our formalism uses an approximation for the probabilities of  contact between the objects.  We show that in moderate pressures the quantities extracted from  the noncentral potential do not rely on its noncentrality, instead we can extract them analytically from a central potential. Our formalism reproduces the nontrivial features of a gas of elongated objects. Finally, we show below a crossover pressure $p_o$ the rotational couplings causes quantities proportional to inverse distance (such as density) are on average deviated from the inverse of average distance.
\end{abstract}
\widetext

\maketitle

\makeatletter

The elasticity of a fluid can be modelled by mechanical response of its rigid boundaries to external deformation forces \cite{landau}. An alternative approach is based on short-range interaction between building blocks  and this has been shown to be consistent with the most of  fluid bulk properties \cite{tonks}. The former  is restricted to  homogeneous fluids, however  the latter is extendible to inhomogeneous case \cite{{takanashi},{ssh}}. Such a discrete approach showed to delineate the elasticity of closely-packed hard objects of perfect spherical symmetry,  \cite{frenkel}. A natural generalization in the same dimension is to consider  elongated objects instead of spheres.  With centers confined on a straight line, elongated objects are compressed from sides, thus carry a coupling between translational and rotational degrees of freedom. This additional coupling  put them in orientational ordered/disordered phases. Lebowitz, Percus and Talbot in Ref. {[}\onlinecite{lebowitz}{]} studied objects with flat and curved elongations. They found out flat objects (such as needles) resist more against external squeezing pressures from sides.   This can be used for liquid crystals to be modelled, (for instance in nematic model \cite{deGennes}.  The order/disorder phases  may also trigger understanding phenomena such as protein wrapping DNA \cite{chu}, percolation transitions \cite{gust}, and jamming transitions, \cite{{lu},{kardar2}}. 

Despite the importance, little has been understood  about the  statistical mechanics of elongated objects even in one dimension.  Murat, Kantor, and Farago \cite{mk} proposed a direct generalization of Lebowitz formalism to study some features in  the elasticity properties hard elongated objects such as their hard stiffness in short-range  central and non-central potentials. The noncentral potentials depend on the relative distance between objects and their individual rotational angles, while central ones depend only on the distance.   Kantor and Kardar  in Ref. {[}\onlinecite{kk09}{]} revisited the problem of noncentral potential and proposed a collection of parameters such as average distance and elasticity coefficient with nontrivial behaviour under external compression. 

In this paper, we show that in moderate pressures the quantities extracted from  the noncentral potential (i.e. eq. (\ref{eq. needle})) do not rely on its noncentrality, instead we can extract them analytically from a central potential at moderate pressure limits. We show this formalism reproduces the nontrivial features of a gas of elongated objects \cite{{kk09},{kardar2}}.  We examine our formalism beyond Kardar and Kantor's needles model on objects with curvature in their contact interfaces and verify numerical results of Ref. {[}\onlinecite{lebowitz}{]}. In the last section we investigate a crossover pressure at which the averaging over distance inverse deviates from the inverse distance law.

\begin{figure}[h]
\includegraphics[width=8cm]{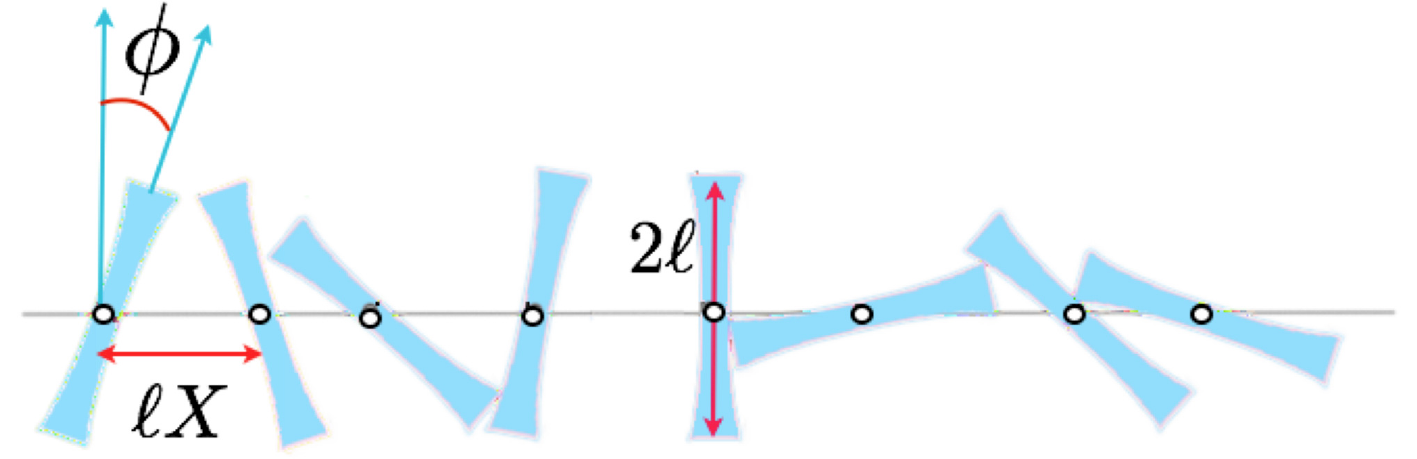} 
  \caption{(Color online) Elongated objects with centers restricted to a straight line. Compressing from sides causes the object to be in contact with each other with centers at distance $\ell X$ and rotational angle $\phi$. }
  \label{fig needles}
  \end{figure}

Consider a large number of identical hard (and needle-like) elongated objects with centers confined to be mobile  on a straight line. The length of each object is $2\ell$. Each object contains two degrees of freedom: A translational $X$, and a rotational $\phi$. We choose the angle reference axis to be perpendicular to the central axis.  Due to the transversal symmetry of elongated objects the rotational angle is $\pi$-periodic, between $-\pi/2$  and $\pi/2$, see Fig. (\ref{fig needles}). 

Applying pressure from the sides will position objects in contact to each other.  The contact distance between  two adjacent centers depend on the individual orientations of the objects and in low pressures could be as large as the the needles length.  Note that the distance between adjacent objects can be greater than the contact length as long as they do not touch upon each other.  We work out our formalism for a general set of objects, yet the picture is similar to the needles model described in Ref. [\onlinecite{lebowitz}] for which the distance between two adjacent objects in contact is  $\ell X_{i,i+1}$ with

\begin{equation}
X_{i,i+1}=\frac{\sin(|\phi_i-\phi_{i+1}|)}{\max \cos
(\phi_i, \phi_{i+1} }.
\label{eq. needle}
\end{equation}

\begin{figure}[h]
\centering
\parbox{3.5cm}{\includegraphics[width=4.5cm]{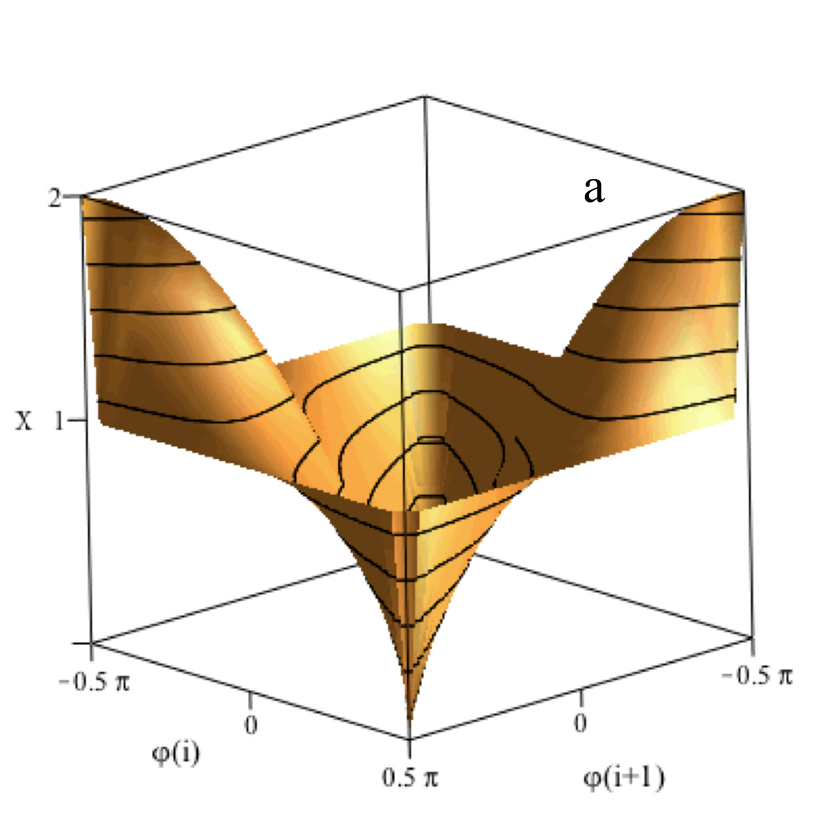}}
\qquad
\begin{minipage}{1.7in}%
{\includegraphics[width=4.5cm]{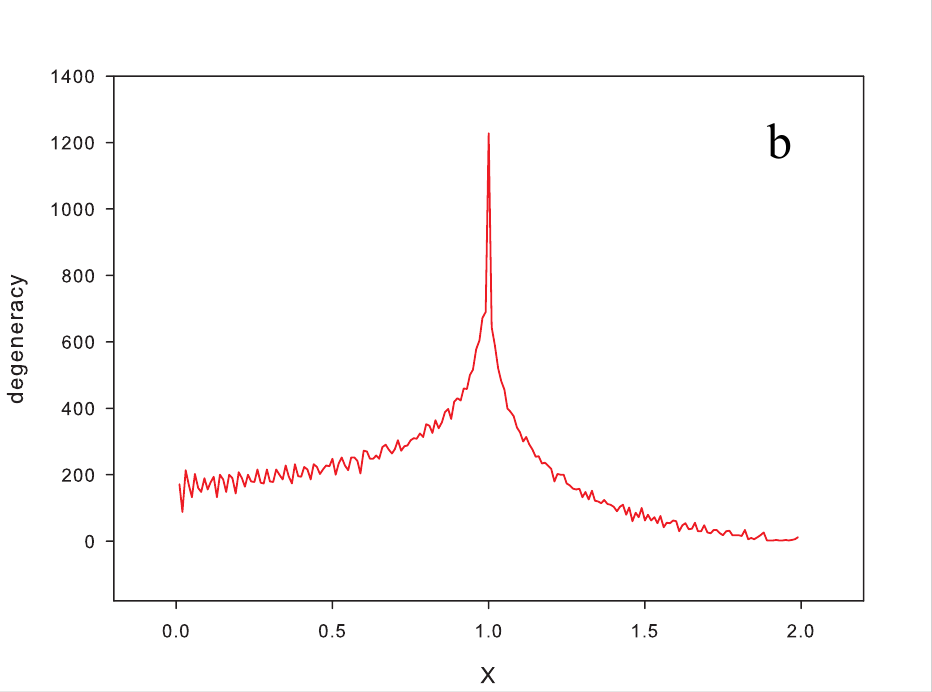}}
\end{minipage}%
\caption{(Color online)  (a) Contact $(\phi,X)$-space in the needles model eq. (\ref{eq. needle}). (b) The degeneracy of $\phi_{i,i+1}$ configurations associated to any length $X$.}%
\label{fig 1}%
\end{figure}

The contact ($X,\phi$)-space diagram is depicted in Fig. (\ref{fig 1}a). There are 2 points with  maximum  distance where $\phi_i=-\phi_{i+1}=\pi/2$.  Also,  minimum distance (i.e. $X=0$)  is at the line $\phi_i = \phi_{i+1}$.  Contours represent configurations of different $\phi_{i, i+1}$ reproducing the same distance and the length of a contour is proportional to the configuration degeneracy. The longer a contour line is the more popular that distance is among configurations.   Figure (\ref{fig 1}b) indicates this degeneracy, to the discreteness of $5\%$ degree, as a function of distance $X$. 

One can see in Fig. (\ref{fig 1}b) the slow variation of degeneracy, except in the vicinity of $X=1$ where the degeneracy blows out;  but why distances about $X=1$ are heavily degenerate? This is mostly due to the pathological behavior of eq. (\ref{eq. needle}) at this point.  The needle $i$  aligned tangentially on the central axis ($\phi_1=\pm \pi/2$) touches upon the centre of its adjacent needles (i.e $i\pm1$). In this case the distance between the two adjacent needles is insensitive to the the orientation of the object $i\pm 1$; thus a large degeneracy is expected.  We regularize the degeneracy by considering a constant discreteness unit for angles and practically  associated a finite ratio  $g_X$ to a degeneracy contour. Regularized degeneracy is defined as $\Omega_X=g_X/g_{X_0}$, where $g_{X_0}$ is a typical finite degeneracy (here we used $X_{0}=0$).  Note that $\Omega(X=0)=1$. The degeneracy fluctuations in  Figure (\ref{fig 1}b) is the artefact of angle discretization process for counting states and can be smoothened by decreasing the angle steps.  Note that discrete regularization make $X$ to carry a finite degeneracy $\Omega(X)$.

Takanashi in Ref. [\onlinecite{takanashi}] assumed  a free energy to `hard' objects by considering  interaction between  a pair of adjacent objects $i$ and $i+1$  by the hard potential $U(x_i, x_{i+1})$.  This potential $U(x_i, x_{i+1})$ is zero at distances greater than the contact distance $X_{i,i+1}$; otherwise it is infinite.  Note that the distance in eq. (\ref{eq. needle}) is only the contact distance and indeed the objects could have distances greater than $X$ which make them not to touch each other; similar to free molecules in a gas.

Consider the system is under a  moderate pressures and exquisitely returns to its equilibrium very rapidly.  This simplifies the system to be unaffected thermally.  Also  we consider  the kinetic energy vanishes at equilibrium.  A number of $N$ needles compressed by the external pressure $P$ carry the Gibbs free energy $G = E - TS + PV$ with the system size $V$, the internal energy $E$, the entropy $S$, and the temperature $T$.  The  partition function is formally written as $Z_G=\sum \exp(-\beta G)$.  A hard potential contribution to the partition function 

\begin{equation}
Z_G = \prod_{i=1}^N \int d\phi_i \int dx_i  e^{-\beta \left(U\left(x_i-x_{i+1}\right)+ P |x_i-x_{i+1}|\right)}.
\end{equation}

  Let us define an independent variable that measures the distance between a pair of adjacent needles; $s_i:=|x_i-x_{i+1}|$. Hard potential energy will simplify the partition function into 
  
  \begin{eqnarray*}
  Z_G &=& \prod_{i=1}^{N} \int d\phi_i \int_{\ell X}^{\infty}  ds_i \exp(-\beta p s_i)\\
  &=&  (\beta p)^{-N}  \prod_{i=1}^N  \int d\phi_i\  \exp(-\beta p\ \ell  X(\phi_i, \phi_{i+1}))
  \end{eqnarray*}

  Note that open chain and closed chain of needles have the same free energies in the large $N$ limit. More precisely one can show the difference between the open ended and closed boundary conditions in the partition free energy is in the end points energy that is the excess of the order $1/N$ more in the open case.  
  
  In statistical mechanics one can show that if the vast  majority of states have slow varying degeneracy while a few   carry exponentially larger (or smaller) degeneracies, average over all states is almost insensitive to the degeneracy of that minority.   This strikes a possible approximations that helps to simplify our problem so that we simply ignore the rise of degeneracy about $X=1$.  Moreover, one can see in Fig. (\ref{fig 1}b) that the degeneracy is a slowly varying function. After ignoring the resonance about $X=1$, one may assume the degeneracy variation follows approximately a linear function of $g(x)\sim -aX+b$ for $a,b >0$; however because physical parameters are proportional to $\ln Z$ the variation of the degeneracy at different $X$ is negligible in the leading order term.  
  
  We simplify the problem into a uniform distribution of degeneracy.  Also we assume the contact distance between any adjacent objects is randomly distributed between 0 and $2\ell$; $X(x)=2x$ where $0 \leq x\leq 1$.  We keep the maximum contact distance between two adjacent particles to be  2 to make our results  comparable with the original needles model, otherwise we can normalize it to one.  In the low pressures limit the object behaves as point particles. In contrast, at high pressure limits when pairs become exceedingly packed, the quantitates extracted from the Gibbs energy  become  highly dependent into the individual angles of objects.   We restrict our method  to moderate pressures, where particles are not densely packed.  The distance of pairs in low to intermediate pressure limits is independent of other pairs, thus the partition function is simplified as a product;  $Z=f^{-N}   Z_1^N$.  Let us define the dimensionless force $f=\beta p \ell$.   The two-body partition function is 
  
\begin{equation} \label{eq. Z1}
  Z_1 = f^{-1} \left( \int_0^1 dx e^{-2f x}\right) =  \frac{1-e^{-2f}}{2f^2}.
\end{equation}

 The bulk partition function of $N$ particles is $Z\sim Z_1^N$, therefore the Gibbs free energy per particles is

\beq \label{gibbs}\beta G/N = - \ln Z_1= 2\ln f - \ln (1-e^{-2f}). \eeq

In small pressure one can expand eq. (\ref{eq. Z1}) in terms of force and approximate the partition function of a pair as $Z_1=(1-f + \frac{2^2}{3!}f^2-\frac{2^3}{4!}f^3+\cdots)/f$. In the leading order becomes $Z_1 \sim 1/f $ and consequently the Gibbs free energy  becomes $\beta G/N \sim \ln f $ that is consistent with the free energy  of point particles.

%

Following the formalism originally proposed for spherically symmetric objects in \cite{ssh} and later on extended to the correlation functions of elongated objects in \cite{mk}, bulk stress can be introduced for a one-dimensional system as the average distance between building blocks.  This quantity can be defined as $a/\ell =  \partial (\beta G/N) / \partial f$, \cite{kk09}. Substituting the Gibbs energy of Eq. (\ref{gibbs}) into this definitions, the interparticle distance becomes: 

\beq \label{eq a}
\frac{a}{\ell} =  \frac{2}{f}-\frac{2}{e^{2f}-1}
 \eeq

By decreasing  pressure the average distance increases. There is no upper bound for the increase of mean distance as net distance between two adjacent objects can be infinitely large due to their lack of contact.  However, once the molecules start not to contact each other their statistics starts to converge to a gas of point particles. In fact there is a minimum onset force $f_o$ where the objects are in touch below which (i.e. $f<f_o$) their angular degrees of freedom become irrelevant to the statistical mechanics.  The onset pressure can easily be calculated by considering that at $f<f_o$ the average distance starts to become greater than $ 2\ell$. As the onset pressure is the solution to the equation  $f_0 [1-\exp(-2f_o)]^{-1}=1$ that after taylor expansion provides the onset pressure to be $f_o=0.8$.

Similarly, the elastic coefficient has been defined in \cite{kk09} by $C=-a/[\partial a/\partial p] +p$.  Rescaling the coefficient into $\beta \ell C$ simplifies this into  $-a/[\partial a/\partial f] +f$. Using the Gibbs energy of Eq. (\ref{gibbs}) an analytical formula for the elastic coefficient becomes

\begin{equation}\label{eq c}
    C = \frac{f}{2}\ \frac{4\cosh(2f)-(2f^2+f+4)+f\exp(-2f)}{\cosh(2f)-(f^2+1)}
\end{equation}

\begin{figure}[h]
\center
\begin{overpic}
[scale=.42,unit=0.5mm]%
{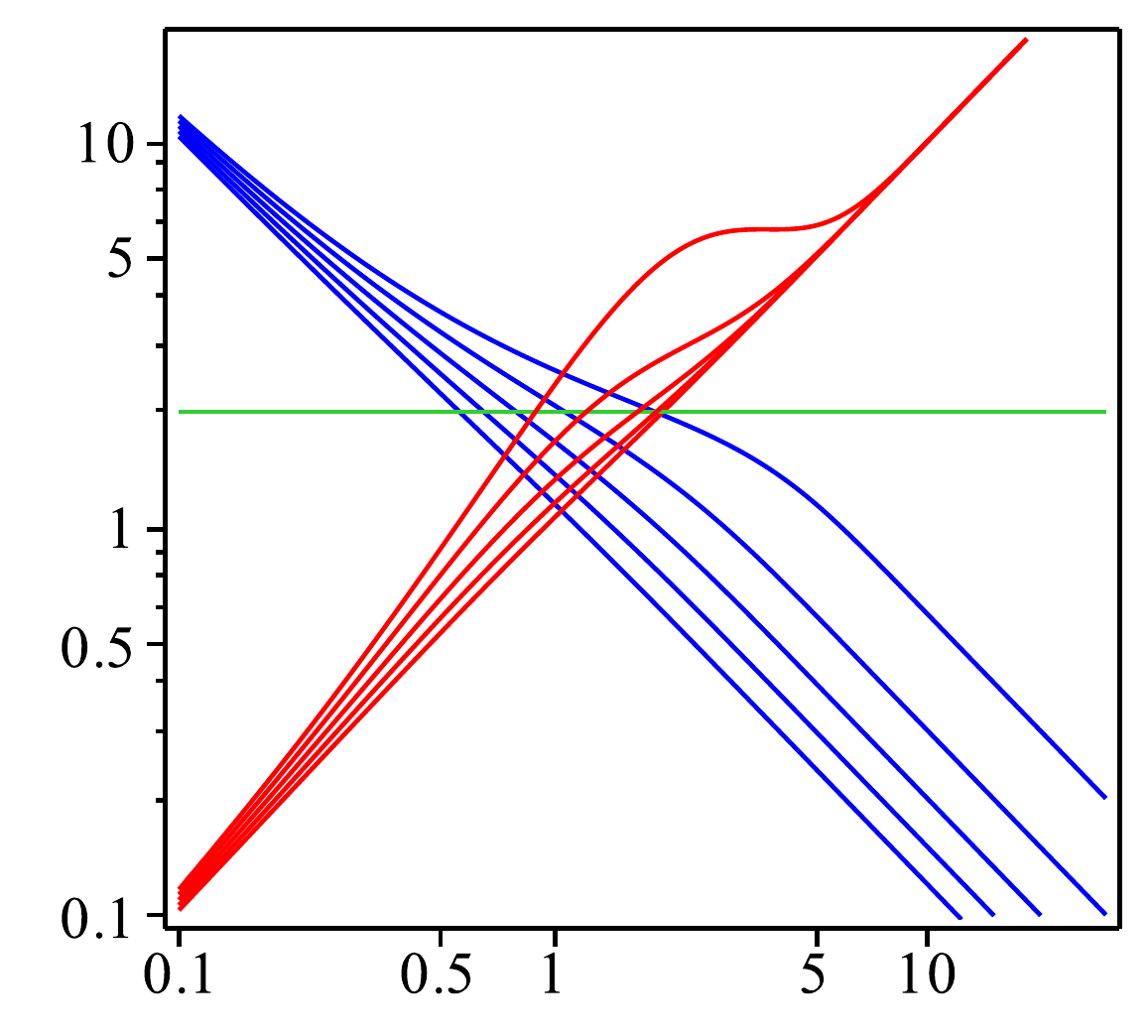}
\put(35,135){\large $a/\ell$}
\put(35,63){\large $\beta \ell C$}
\put(160,7){\large $f$}
\put(121,77){\includegraphics[scale=.1]%
{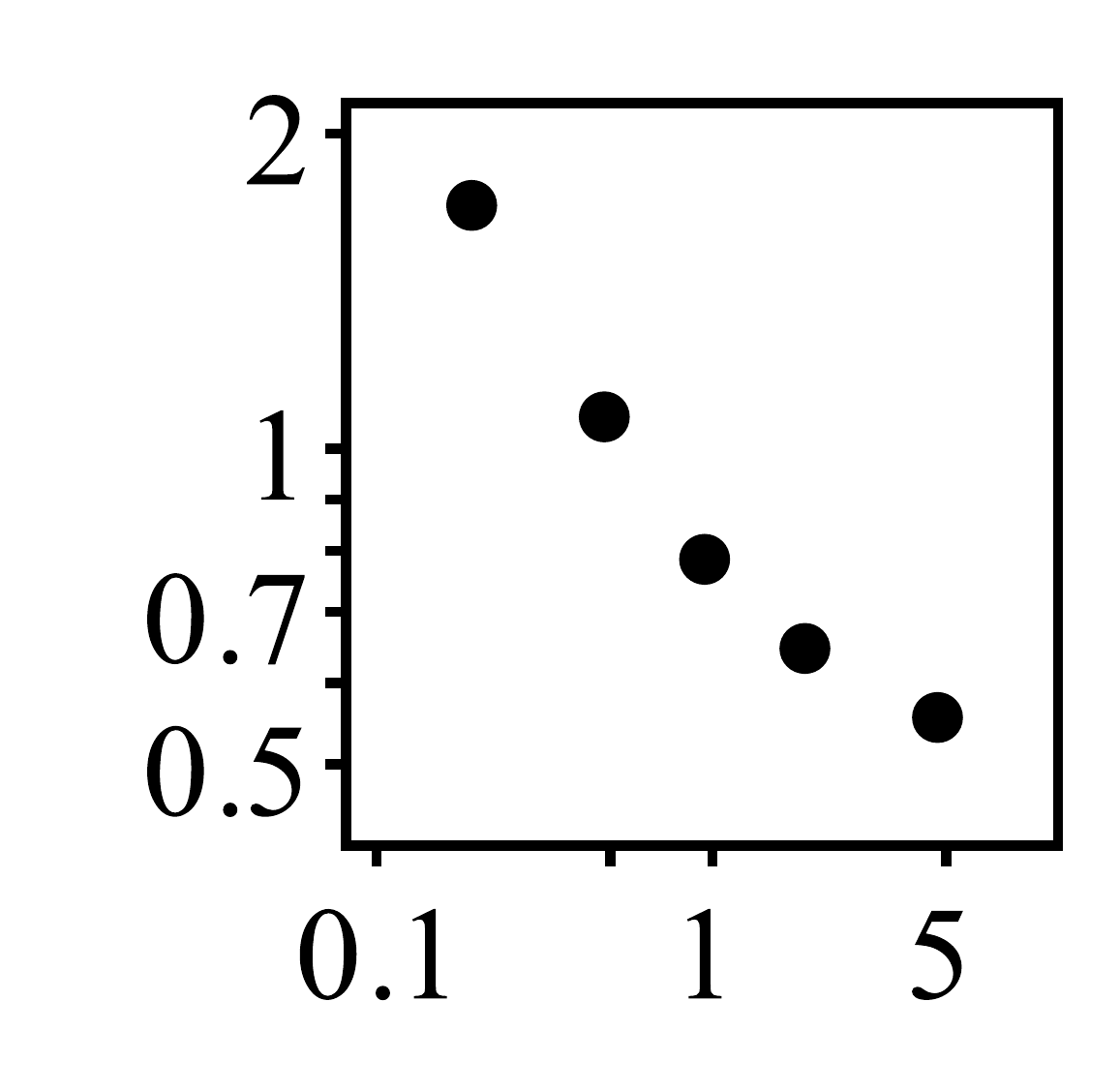}}
\put(140,76){$\alpha$}
\put(123,103){$f_o$}
\end{overpic}
\caption{(Color online)  Average distance $a/\ell$ and the elasticity coefficient $\beta \ell C$.  From   top to bottom $\alpha=0.2,\ 0.5,\ 1,\ 2,$ and 5, respectively. The green line indicates the onset above which the gas effectively behaves similar to hot gas of point particles.}
\label{fig 4}
\end{figure}
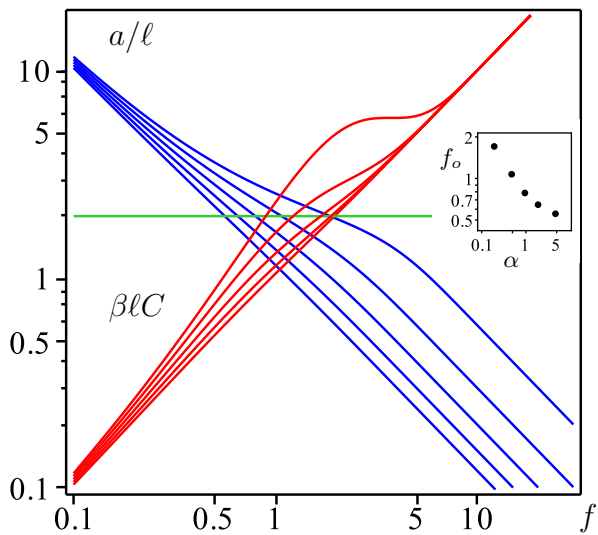

We plotted the average distance $a$ and the elasticity coefficient $C$  labelled as $\alpha=1$ in Fig.  (\ref{fig 4}). In this figure the negative- and positive-slope curves indicate average distances and elasticity coefficients, respectively.  The dash lines denote the results for point particles. The green line indicates the $2\ell$ above which the average distance between adjacent particles become greater than the maximum contact distance. However, this onset pressure is not expected to be a critical pressure. To understand this recall that it  is not necessary for two adjacent particles to exceed the distance $2\ell$ in order to lose their contact; Instead, there is a possibility their distance become greater than $2x$.

One can check that our analytical formulas for average distance and elasticity coefficient are in agreement with the corresponding numerical solutions of exact model summarized in the Fig. (4) of Ref. {[}\onlinecite{kk09}{]}.

Let us now step beyond the limitations of needles model and use our approximation logic as a tool to predict similar behaviour of other types of objects. Lebowitz in Ref. {[}\onlinecite{lebowitz}{]} proposed to deforme distance in curved interface objects to $x^\alpha$, however because we are interested to compare results of different exponent $\alpha$ we keep the distances to be normalized to 2; thus 

\begin{equation}
\label{eq gamma d}
    X_{i, i+1}(x)=2x^\alpha.
\end{equation}

Straight interfaces (such as needles)  are characterized with $\alpha=1$. Convex objects with $\alpha>1$. A special case is ellipsoids for which $\alpha=2$. Concave objects are also characterized by $\alpha<1$.   In order to avoid complications of a preferred orientation let us consider the object widths are negligible. This immediately imposes a restriction on the accuracy of our forthcoming analysis that the interface curvature should be only slightly curved from flatness.   

The partition function per particle using eq. (\ref{eq gamma d}) is

\begin{eqnarray}\label{eq Z gamma}
  Z_1 &=& \frac{1}{f} \int_0^1 e^{-2 f x^\alpha}dx \nonumber \\
   &=& \frac{1}{ 2^{1/\alpha} \alpha \ f^{1+1/\alpha}} \left( \Gamma\left( \alpha^{-1}
   \right)- \Gamma\left( \alpha^{-1}, 2f\right)\right),
\end{eqnarray}
where $\Gamma(a,x) = \int_x^\infty t^{a-1} e^{-t} dt$. The average  distance for different choices of $\alpha$ can also be derived,

\begin{equation}
\label{eq a gamma}
    \frac{a}{\ell}= \frac{1+1/\alpha}{f} - \frac{2^{1/\alpha}f^{1/\alpha-1}\exp({-2 f})}{\Gamma(\alpha^{-1})-\Gamma(\alpha^{-1}, 2f)}.
\end{equation}

To arrive at  this relation we used the chain rule $d\Gamma(a,u(x))/dx = [d \Gamma(a,u)/du]  (du/dx)$, and $d\Gamma(a,x)/dx= - x^{a-1} \exp(-x)$. 

From the partition function, finding the elasticity coefficient from $a$ is straightforward.  The elasticity coefficient $C$ and the aveareg distance $a$ are plotted in the Figure (\ref{fig 4}) for the exponents $\alpha=$0.2, 0.5, 1, 2, 5 from top to bottom, respectively. The mean distance and the elasticity coefficient lines are plotted with negative and positive slopes, respectively.  These results recently have been derived using numerical analysis in \cite{kardar2}.

With the decrease  of $\alpha$ the average  distance between objects and the object  elasticity coefficient starts to deviated from the power law associated to spherically symmetric particles.  The deformation of the average distance  from power law was earlier indicted in Lebowitz et al. \cite{lebowitz}. Our model extend this study to the elasticity coefficient.  Moreover, both of these two quantities for the case of $\alpha=1$ have been studied in Ref. {[}\onlinecite{kk09}{]} and are consistent with our results. As one can see in Fig. (\ref{fig 4}) the smaller $\alpha$ is, the larger the average distance and elasticity coefficient becomes in the range of intermediate pressures.  This is so because the more two adjacent objects are convex on their contact interface, the farther their centers will be in different configurations. 

The green line in Fig. (\ref{fig 4}) indicates the threshold of $a=2 \ell$. At $a>\ell$ the objects start to detach from one another (similar to a hot gas) and the average distance becomes less sensitive to the orientations and behave more like point particle gas.  The inset indicates the  onset pressure $f_o$  associated to $\alpha$. One can see by the increase of $\alpha$ the onset pressure decreases in almost a power law $f_0\sim \alpha^{-1/2}$.

In the last part, we study the average of quantities such as number density of objects  that are scaling  inversely with the distance.  For  point particles, as well as spherically symmetric ones, one may expect the average of inverse distance $\langle 1/r\rangle$ to be equivalent to the inverse of average distance; $1/\langle r\rangle$. However, this is nontrivial to be valid in the presence of rotational degrees of freedom such as in needles model. Below we  derive the ensemble average of density taking into account the rotational couplings. 

In a chain of $N$ elongated objects confined on a line the number density is defined as $\rho_n=\sum_{\textup{Ensemble}} N/L $ where $L$ is the maximum distance between the first and last needle.    One may consider that averaging over the maximum distance in different ensembles may give rise to the approximate formula that the density of needles is $\rho_n=1/\langle X\rangle_n $.  However, our analysis of the more accurate definition $\rho_n=\langle 1/X \rangle_n$ show that the density is  different at low pressure limit. 

The average value of inverse distance can be calculated in the followings:
\begin{equation} 
\ai_n= \frac{\int dx \frac{1}{X(x)}\ e^{-fX(x)}}{\int dx\ e^{-fX(x)}}
\end{equation} 

One can rewrite this equation by taking the derivative from both sides with respect to $f$ 

\beq
\frac{d }{d f}\ai_n= -1 + \ai_n a \label{eq dainv/df}
\eeq

From eq. (\ref{eq dainv/df}) one can see the equality between $\ai$ as $1/a$ is valid only if the left hand side of eq. (\ref{eq dainv/df}) becomes zero.  However, this is not the case in general because as one can see in Fig. (\ref{fig 4}) the minimum contact distance $a$ between particles depends strictly on pressure. 

Substituting the average distance $a$ from eq. (\ref{eq a}) in eq. (\ref{eq dainv/df}) gives rise to the following equation:
\beq
\frac{d}{df}\ai_n =  -1 + \left(\frac{2}{f}-\frac{2}{e^{2f}-1}\right) \ai_n. \label{eq.ainverse}
\eeq

Solving this equation can provide us with the explicit dependence of the average potential to the pressure.  This differential equation has a solution for the needles density $\rho_n=\ai_n = f +2f^2\ E_1(2f) [1-e^{-2f}]^{-1}$, where $E_1(z) = \int_1^{\infty}\exp(-tz)t^{-a} dt$ is an exponential integral function. $E_1(z)$ is known to  exponentially suppress at large $f$ and scaling logarithmically with $1/f$ for small $f$.  Usually this function can be approximate  by one of its bracketing function bounds (e.g. see \cite{abr}). In our problem, this allows to replace it with $\frac{1}{2}\exp(-2f)\ln(1+1/f)$. Consequently, the expectation value of inverse distance becomes:
\beq
\rho_n=\ai_n = f +\frac{f^2 \ln(1+\frac{1}{f})}{e^{2f}-1} \label{Eq. aEi}
\eeq

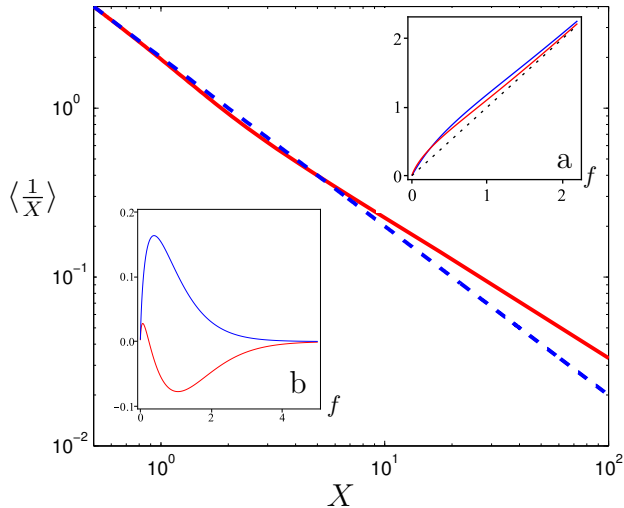
\begin{figure}[h]
\center
\begin{overpic}
[scale=.73,unit=0.5mm]%
{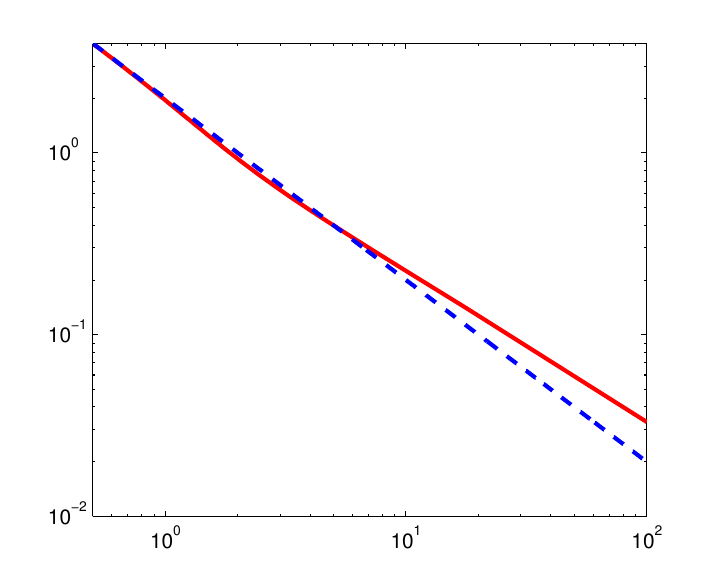}
\put(85,0){\large $X$}
\put(0,80){\large $\langle \frac{1}{X}\rangle $}
\put(24,17){\includegraphics[scale=.155]%
{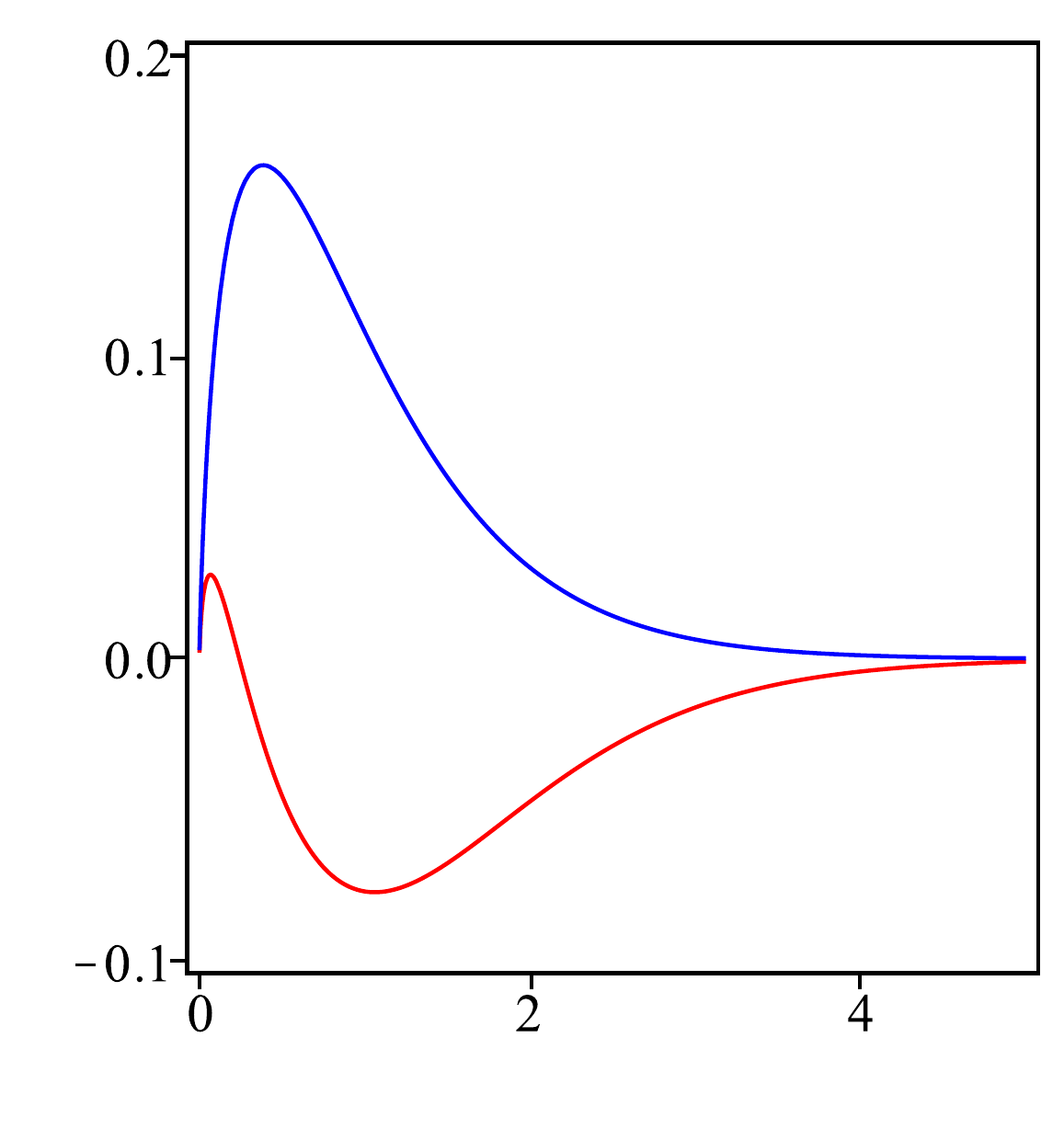}}
\put(75,30){\large b}
\put(98,78){\includegraphics[scale=.145]%
{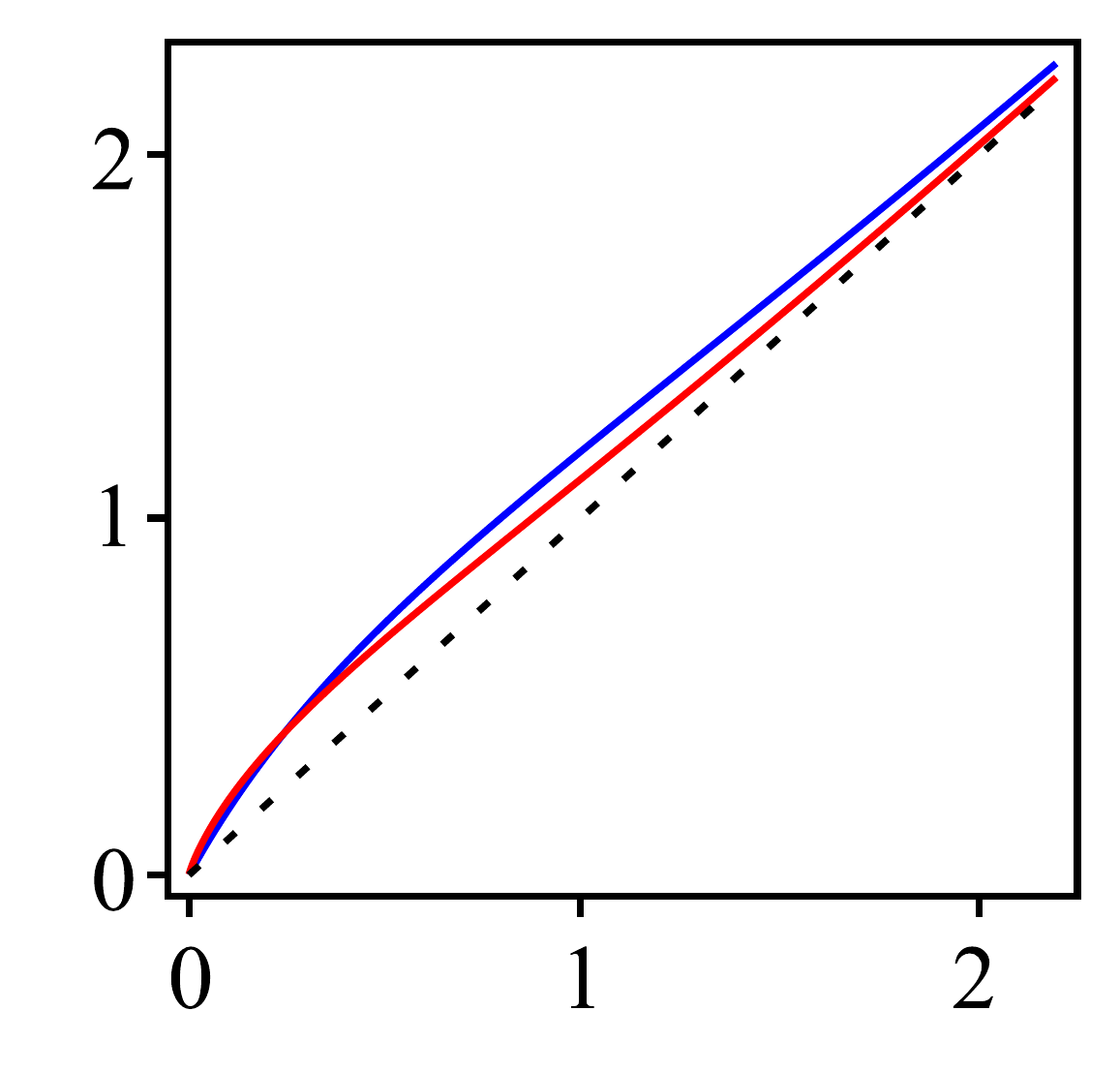}}
\put(146,90){\large a}
\put(85,25){$f$}
\put(153,86){$f$}
\end{overpic}
\caption{(Color online) The distance variation of ensemble average density in needles model and its comparison with the inverse distance (dash). Insets:  (a) The inverse of average distance in needles $1/\langle X\rangle_n$ (top),  average density $\langle 1/X\rangle_n$ in needles (middle), average density i spherically symmetric objects (bottom dashed). (b) $\langle 1/X \rangle_n- \langle 1/X\rangle_{ss}$ (top), which has the information of deviation of density in needles compared to spherically symmetric objects, and $\langle 1/X \rangle_n- 1/ \langle X\rangle_{n}$ (bottom) which indicates the deviation of the ensemble average density from the inverse of average distance. }
\label{fig 5}
\end{figure}

The first term in the right side of eq. (\ref{Eq. aEi}) is the spherically symmetric (ss)  leading order in small pressure limit.  In this limitthe density follows the inverse law   $\rho_{ss} = \ai_{ss} = f = 2/\langle X\rangle_{ss}$.  Note from eq. (\ref{eq a}) that in spherically symmetric case $\langle X \rangle_{ss} = 2/f$.  However, the second term in eq. (\ref{Eq. aEi})  is the explicit contribution of elongation to the density.  

Figure (\ref{fig 5}) indicates density as a function of distance and compare it with inverse distance (dash lie). One can see that at low distance (i.e. high pressure) they are very close to each other; however as the distance between objects increases the particle density deviates from $1/X$. This can be understood from Fig. (\ref{fig 5}a) by comparing  $1/\langle X\rangle_n$ (top line) in needles, $\langle 1/X \rangle_n$ (middle line) in needles, and $\langle 1/X\rangle_{ss}$ (dotted line) in spherically symmetric model as functions of external force.   At intermediate pressures  one can see the statistical average of density in needles model is greater than that in spherically symmetric objects and point particles; however, it is smaller than $1/\langle X\rangle_n$.

Fig. (\ref{fig 5}b) presents $\langle 1/X \rangle_n- \langle 1/X\rangle_{ss}$ (upper line)  and $\langle 1/X \rangle_n-  1/\langle X\rangle_n$ (lower line). As a result by decreasing pressure the density of needles exceeds that in balls almost exponentially.  At extremely low pressure these two becomes closer as it is expected for the situation of very low density gas. Moreover,  one cannot rely on the quantity $1/\langle X\rangle_n$ too to represent the density as it is larger than the actual value. The true ensemble density of needles is greater than that in spherically symmetric and point particle gases and this error can be up to 30$\%$ at low pressures. Moreover, there is also a noticeable difference between the density and the inverse of the average distance. The true density could be  up to 18$\%$ \emph{smaller} than inverse of mean distance.

In conclusion:  We presented an analytical solution for  the gas of needles model that interpolates between the low and high pressure behaviour. We  extended our solution to the cases of elongated objects other than needles and reproduced the average distance and elasticity coefficient in consistency with numerical results taken from the exact models.  In our approximation we eliminate the dependence of free energy in the adjacent individual angles.  Although we expected this approximation alters the order-related quantities  at high pressures, where the objects become highly ordered;  interestingly we noticed the collective quantities that we study here are unrelated to the ordering and could be described without noncentrality of potentials.  We derived the ensemble density of needles and found out at intermediate pressures it does not follow up an inverse distance law.  

This model serves as the initial point to the study of statistics of elongated objects analytically and  may also serve to extend the problem in higher dimensions. 

\section*{Acknowledgement} This work was supported by NSERC Canada.


\end{document}